# Power System Supplementary Damping Controllers in the Presence of Saturation

M. Ehsan Raoufat, *Student Member, IEEE,* Kevin Tomsovic, *Fellow, IEEE,* Seddik M. Djouadi, *Member, IEEE*

*Abstract*—This paper presents the analysis and a method to design supplementary damping controllers (SDCs) for synchronous generators considering the effects of saturation limits. Usually such saturations of control signals are imposed in order to enforce practical limitations such as component ratings. However, to guarantee the stability in the presence of saturation limits, the state trajectories must remain inside the domain of attraction (DA). In this paper, the domain of attraction of a single-machine infinite-bus (SMIB) power system with saturation nonlinearity is estimated and compared with the exact description of the null controllable region. Then, state-feedback controllers are designed to enlarge the DA. Our analysis shows that nonlinear effects of saturation should be considered to guarantee stability and satisfactory performance. Simulation results on a detailed nonlinear model of a synchronous generator indicate that the DA enlarges with the proposed controller. The results also indicate that Critical Clearing Time (CCT) and damping of the system with saturation can be improved by the proposed method.

## NOMENCLATURE

| | |
|---|---|
| $\delta$ | generator angle, in rad; |
| $\omega_s$ | synchronous and speed, in rad/s; |
| $\omega_r$ | normalized speed, in pu; |
| $H$ | inertia constant, in s; |
| $T_M$ | mechanical torque, in pu; |
| $E'_q$ | q-axis transient voltage, in pu; |
| $I_d, I_q$ | d-axis and q-axis current, in pu; |
| $D$ | generator damping coefficient; in pu; |
| $T'_{d0}$ | d-axis time constant, in s; |
| $X_l, X_t$ | line and transformer reactance, in pu; |
| $X_d, X_q$ | d-axis and q-axis reactance of the generator, in pu; |
| $X'_d$ | d-axis transient reactance of the generator, in pu; |
| $E_{fd}$ | field voltage, in pu; |
| $V_{ref}$ | reference voltage, in pu; |
| $V_t$ | generator terminal voltage, in pu; |
| $V_d, V_q$ | d-axis and q-axis voltage, in pu; |
| $V_\infty$ | infinite bus voltage, in pu; |
| $V_s$ | supplementary control input, in pu; |
| $K_A$ | excitation gain; |
| $T_A$ | excitation time constant, in s; |
| $K_D$ | damping coefficient, in pu/(rad/s); |
| $K_S$ | synchronizing coefficient, in pu/rad; |

This work was supported in part by the National Science Foundation under grant No CNS-1239366, and in part by the Engineering Research Center Program of the National Science Foundation and the Department of Energy under NSF Award Number EEC-1041877 and the CURENT Industry Partnership Program.

M. Ehsan Raoufat, Kevin Tomsovic and Seddik M. Djouadi are with the Min H. Kao Department of Electrical Engineering and Computer Science, The University of Tennessee, Knoxville, TN 37996 USA (e-mail: mraoufat@utk.edu).

978-1-5090-5550-0/17/$31.00 ©2017 IEEE

## I. INTRODUCTION

One of the most common instability problems in power system is low frequency electromechanical oscillations, which may grow and result in loss of synchronization under certain conditions [1]. Supplementary damping controllers (SDCs) based on wide-area measurements [2] can be used to damp these oscillations. Multiple approaches have been proposed in literature to design SDCs for power system components, including traditional synchronous generators [3]–[6], modern FACTS devices [7], [8], energy storage systems [9] and renewable resources [10]. These efforts generally do not consider the nonlinear effects of hard saturation limits on control signals. Moreover, new generation sources connected to the grid through inverters, such as, photovoltaics, have the ability to provide damping signals but only within a narrow range dependent on operating conditions. It is critical to consider these actuator constraints for such components. Moreover, the actuator saturation effect is also a common phenomenon in other fields including saturation of robotic actuators or modulation signals of power converters [11]–[14].

In this work, saturation, or a hard limit, is considered for the control signals. Note this is different from the traditional magnetic saturation of generators but instead reflects the practical limitations of equipment ratings and can be expressed using hard saturation limits restricting the amplitude of the controller output. These limits can be considered in the excitation to prevent undesirable tripping initiated by over-excitation or under-excitation of generators [15]. In case of generator SDCs, saturation limits should be considered in the supplementary control input signal and are usually in the range of $\pm 0.05$ to $\pm 0.1$ per unit which guarantee a modest level of contribution [16]. These limits allow an acceptable control range to provide adequate damping while preventing tripping of the equipment protection. Moreover, this may minimize the negative effects of SDCs on the voltage regulatory response.

There exists a large body of work in the control literature on stability analysis of systems with input constraints [17]–[20]. However, the effects of saturation have not been taken into account in previous works for analyzing and designing SDCs [3]–[9]. This paper also extends the work reported in [21] and [22] in which the nonlinear effects of saturation on stability has not been considered. Saturation can negatively impact the performance of SDC since this restriction can limit the control effort available to damp the oscillation and consequently decreases the damping or leads to instability.

The main goal in this paper is to propose a new method to design SDCs which results in a larger domain of attraction (DA) in the presence of saturation. In this paper, the DA of a single-machine infinite-bus (SMIB) power system with saturation nonlinearity is estimated to guarantee a safety region of initial conditions (which may be caused by faults) and compared with exact description of the null controllable region. Then, state-feedback controllers are designed to enlarge the DA. In this way, the stability of the SMIB power system is guaranteed and saturation limits are represented in the analysis and design procedure. Our analysis shows that nonlinear effects of saturation should be considered to guarantee the stability and satisfactory performance. Moreover, enlargement of DA effectively enhances the Critical Clearing Time (CCT) and damping of the system with saturation can be improved by the proposed method.

The rest of this paper is structured as follows: preliminaries on dynamic modeling of the SMIB power system with input saturation nonlinearity are described in Section II. Section III is devoted to optimal estimation of the DA for a power system with pre-designed state-feedback controller. For comparison purposes, exact description of the null controllable region of the SMIB power system is given in this section. An optimization method to design state-feedback controllers and to enlarge the DA is described in section IV. The results are compared through detailed nonlinear simulations. Concluding remarks are presented in section V.

## II. Dynamic Model of Power System

In this study, a SMIB power system model is considered. However, the analysis can be extended to cases of SMIB power system with FACTS devices. As shown in Fig. 1, this system consists of a synchronous generator connected through two transmission lines to an infinite bus that represents an approximation of a large system. A flux-decay model of the synchronous generator equipped with a fast excitation system can be represented by the following set of dynamic equations:

$$\dot{\delta} = \omega_s(\omega_r - 1) \tag{1}$$

$$\dot{\omega}_r = \frac{1}{2H}\left[T_M - \left(E'_q I_q + (X_q - X'_d)I_d I_q + D\omega_s(\omega_r - 1)\right)\right] \tag{2}$$

$$\dot{E}'_q = -\frac{1}{T'_{d0}}\left[E'_q + (X_d - X'_d)I_d - E_{fd}\right] \tag{3}$$

$$\dot{E}_{fd} = -\frac{E_{fd}}{T_A} + \frac{K_A}{T_A}\left[V_{ref} - V_t + \text{sat}(V_s)\right] \tag{4}$$

while satisfying the following algebraic equations:

$$R_e I_q + X_e I_d - V_q + V_\infty \cos(\delta) = 0 \tag{5}$$

$$R_e I_d - X_e I_q - V_d + V_\infty \sin(\delta) = 0 \tag{6}$$

$$V_t = \sqrt{V_d^2 + V_q^2} \tag{7}$$

where $R_e$ and $X_e = X_t + \frac{1}{2}X_l$ are the total external resistance and reactance, respectively. One of the nonlinearities associated with the above model is due to the hard saturation limit considered on the supplementary control input of the exciter. In this work, the hard limit is defined as:

$$\text{sat}(V_s) = \text{sign}(V_s)\min\{m, |V_s|\} \tag{8}$$

$$m = V_s^{\max} = -V_s^{\min} \tag{9}$$

where the nonlinearity is assumed to be symmetric and $\pm m$ are the saturation limits. To design a SDC and study the effects of saturation, the above nonlinear model can be linearized around the nominal operating point and expressed in the following state-space representation:

$$\dot{x}(t) = Ax(t) + B\text{sat}(V_s) \tag{10}$$

where

$$x = \begin{bmatrix} \Delta\delta & \Delta\omega_r & \Delta E'_q & \Delta E'_{fd} \end{bmatrix}^T$$

$$A = \begin{bmatrix} 0 & \omega_s & 0 & 0 \\ -\frac{K_1}{2H} & -\frac{D\omega_s}{2H} & -\frac{K_2}{2H} & 0 \\ -\frac{K_4}{T'_{d0}} & 0 & -\frac{1}{K_3 T'_{d0}} & \frac{1}{T'_{d0}} \\ -\frac{K_A K_5}{T_A} & 0 & -\frac{K_A K_6}{T_A} & -\frac{1}{T_A} \end{bmatrix}, B = \begin{bmatrix} 0 \\ 0 \\ 0 \\ \frac{K_A}{T_A} \end{bmatrix} \tag{11}$$

and $K_1$–$K_6$ are the well-known linearization constants presented in Appendix A [23].

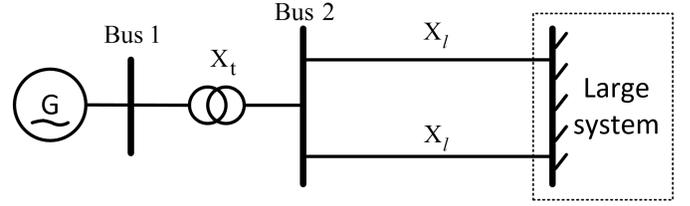

Fig. 1. A single-machine infinite-bus power system.

Although the large gain of the excitation system $K_A$ can reduce the generator terminal voltage fluctuations, it can also introduce negative damping torque to the system at times sufficient to result in instability. To increase the guaranteed region of stability, the DA for this unstable system with saturated feedback controller should be optimized.

## III. Estimation of the Domain of Attraction

In this section, we address the problem of estimating the DA for a system with actuator constraint and a pre-designed state-feedback law. Consider the system of equations (10) with unstable matrix $A \in \Re^{n \times n}$ and state-feedback control law defined by $V_s = Fx(t)$, the closed loop system can be expressed as follows:

$$\dot{x}(t) = Ax(t) + B\text{sat}(Fx(t)) \tag{12}$$

where the DA with the above transition map of $\phi : (t, x_0) \to x(t)$ can be defined as

$$\mathcal{D} := \{x_0 \in \Re^n : \lim_{t \to +\infty} \phi(t, x_0) = 0\} \tag{13}$$

Without saturation limits, the DA for stable $A + BF$ is $\Re^n$; however in the presence of saturation, DA is a subset of

$\Re^n$ and needs to be estimated. There are various methods to approximate the DA [17], [19]. In this paper, we follow the work in [20] to obtain the least conservative estimation based on the Lyapunov function. For a matrix $P > 0$ and $\eta > 0$, we can define the ellipsoid $\varepsilon(P, \eta)$ representing the DA as follows:

$$\varepsilon(P, \eta) = \{x \in \Re^n, \quad x'Px \leq \eta\} \quad (14)$$

which is a contractive invariant set inside the DA. Each eigenvalue of $P$ is related with the length of one axis. Since $\text{trace}(P)$ is the sum of its eigenvalues, its minimization leads to the largest ellipsoid having the same weight in all directions. This problem can be formulated indirectly as the following optimization.

$$\min_{S,W,Z,M_w} \text{trace}(M_W)$$

$$\text{subject to} \quad \begin{bmatrix} M_W & I_n \\ I_n & W \end{bmatrix} > 0$$

$$\begin{bmatrix} W(A+BF)' + (A+BF)W & BS - Z' \\ SB' - Z & -2S \end{bmatrix} < 0$$

$$\begin{bmatrix} W & WF' - Z' \\ FW - Z & m^2 \end{bmatrix} > 0 \quad (15)$$

where $S$, $W$ and $M_W$ are symmetric positive definite matrices and the ellipsoid $\varepsilon(P, \eta)$ with $P = W^{-1}$ is the estimated DA. In the above optimization, minimizing $\text{trace}(M_W)$ implies the minimization of $\text{trace}(P)$ as the first constraint guarantees that $P < M_W$. The second and third constraints guarantee asymptotic stability of saturated system via a quadratic Lyapunov function. Other size criteria such as maximization of the volume or other geometric characterization can also be considered. The estimated result can be compared with null controllable region $\mathcal{C}$, which is defined as the region where there exists an admissible bounded control that can steer the system towards the origin. The null controllable region of an unstable system can be found using the following theorem [24].

**Theorem:** Consider the open loop system (10) with unstable matrix A and B that can be partitioned as follows:

$$A = \begin{bmatrix} A_1 & 0 \\ 0 & A_2 \end{bmatrix}, \quad B = \begin{bmatrix} B_1 \\ B_2 \end{bmatrix} \quad (16)$$

where $A_1 \in \Re^{n_1 \times n_1}$ is semi-stable and $A_2 \in \Re^{n_2 \times n_2}$ is an unstable subsystem. Then null controllable region of the system can be written as:

$$\mathcal{C} = \Re^{n_1} \times \mathcal{C}_2 \quad (17)$$

where $\mathcal{C}_2$ is the null controllable region of the unstable subsystem. Different cases can be considered to find the boundary of $\mathcal{C}_2$; however, in the case of second order subsystems where $A_2$ has a pair of unstable complex eigenvalues $+\alpha \pm j\beta$, $\mathcal{C}_2$ can be characterized as follows:

$$\partial \mathcal{C}_2 = \left\{ \pm \left[ e^{-A_2 t}(I + e^{-A_2 T_p})^{-1}(I - e^{-A_2 T_p}) \right. \right.$$
$$\left. \left. - (I - e^{-A_2 t}) \right] m A_2^{-1} B_2 : t \in [0, T_p) \right\} \quad (18)$$

where $T_p = \frac{\pi}{\beta}$ and $\partial \mathcal{C}_2$ is the boundary of the null controllable region of the second subsystem.

**Example:** Throughout this paper, the SMIB power system is considered to demonstrate the idea and verify the resulting improvement. Parameters of the machine, excitation system, transformer and transmission lines are:

$$X_t = 0.1, \quad X_l = 0.8, \quad R_e = 0, \quad V_\infty = 1.05 \angle 0°,$$
$$X_d = 2.5, \quad X_q = 2.1, \quad X'_d = 0.39, \quad V_t = 1 \angle 15°,$$
$$T'_{d0} = 9.6, \quad H = 3.2, \quad D = 0, \quad \omega_s = 377,$$
$$T_A = 0.02, \quad K_A = 100, \quad V_s^{\max} = -V_s^{\min} = 0.05,$$

Eigenvalue analysis shows that the open loop system has unstable complex eigenvalues of $+0.2423 \pm 7.6064i$ with frequency of 1.21 Hz and damping of $-3.18\%$. The fault considered in this paper is a three-phase fault to ground at bus 2 which is applied at $t = 0.1s$ with fault duration of $t_f$ and cleared without line tripping. The CCT can be defined as the maximum allowable time to clear the fault such that system remains stable and gives information regarding the fault ride-through capability of the generator. First without considering the saturation, a state-feedback controller is designed using the LQR method [25] to minimize the following quadratic performance index:

$$J = \frac{1}{2} \int_0^\infty (x^T Q x + u^T R u) dt \quad (19)$$

where $Q$ is a positive semi-definite weighted matrix related to state cost and $R$ is a positive definite weighted matrix related to the control cost. The parameters of the LQR controller are chosen to be $Q = I_{4 \times 4}$ and $R = 0.1$ and the controller gain can be obtained as follows:

$$F_{LQR} = \begin{bmatrix} -0.7047 & 9.4825 & -3.9325 & -3.1523 \end{bmatrix} \quad (20)$$

In the practice, these controllers can be implemented based on dynamic state estimation using PMU measurements [26]–[29]. The largest guaranteed DA of the SMIB system with the above LQR controller can now be estimated from (15). The ellipsoidal DA is given as $\mathcal{D}_{\mathcal{LQR}} = \varepsilon(P, 1)$ with:

$$P = \begin{bmatrix} 1.1072 & -0.0679 & 1.2368 & 2.5003 \times 10^{-4} \\ -0.0679 & 2844.6 & -46.7344 & -0.0036 \\ 1.2368 & -46.7344 & 19.6725 & 0.0014 \\ 2.5 \times 10^{-4} & -0.0036 & 0.0014 & 9.4678 \times 10^{-4} \end{bmatrix} \quad (21)$$

Using the canonical state-space transformation $z = Tx$ [30], the linear system of equation (10) can be transformed to canonical form and then partitioned as stable and unstable parts similar to (16). Fig. 2 compares cuts of the guaranteed DA of the closed loop system with LQR controller and the null controllable region in the presence of saturation. Using nonlinear simulations, the system with LQR controller has a $CCT$ of 0.081 s. Fig. 3 shows the estimated DA and the extremal trajectory of the nonlinear SMIB system, which demonstrates the accuracy of estimated DA. In the next section, the optimization problem will be modified to design

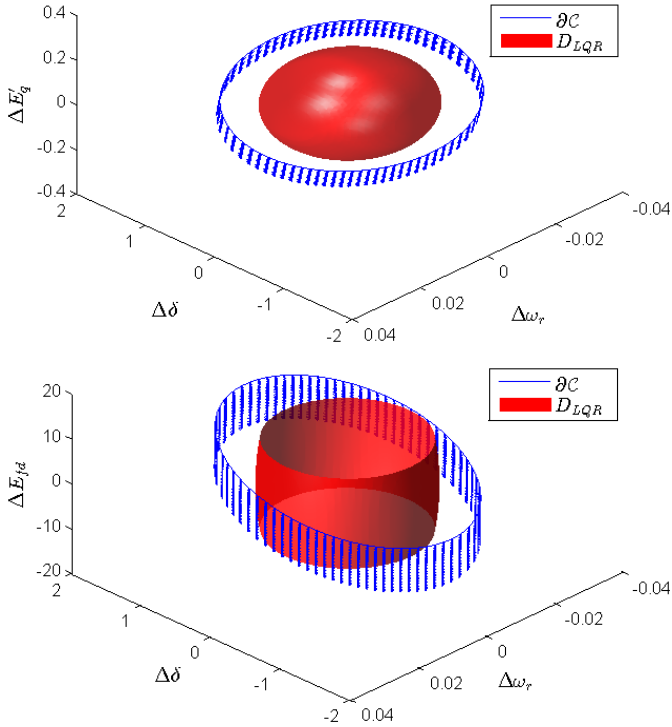

Fig. 2. Boundary of null controllable region ($\partial\mathcal{C}$) and estimated DA for LQR controller ($D_{LQR}$) in the presence of saturation.

a state-feedback controller to expand the domain of attraction toward the null controllable region.

## IV. ENLARGING THE DOMAIN OF ATTRACTION USING STATE-FEEDBACK CONTROLLER

In general, size of the DA depends on the feedback controller and the system constraints, such as, saturation limits. Consequently, the choice of an optimization criterion should include the controller design $F$ in order to enlarge the guaranteed DA. This formulation will introduce bilinear terms of a variable associated with the quadratic Lyapunov function $W = P^{-1}$ and the controller matrix $F$. By introducing an axillary variable $Y = FW$, this problem can be transformed into an LMI problem, namely:

$$\begin{aligned}
\min_{S,W,Y,Z,M_W} \quad & \text{trace}(M_W) \\
\text{subject to} \quad & \begin{bmatrix} M_W & I_n \\ I_n & W \end{bmatrix} > 0 \\
& \begin{bmatrix} WA' + AW + BY + Y'B' & BS - Z' \\ SB' - Z & -2S \end{bmatrix} < 0 \\
& \begin{bmatrix} W & Y' - Z' \\ Y - Z & m^2 \end{bmatrix} > 0 \\
& WA' + AW + (Y'\Gamma_j^+ + Z'\Gamma_j^-)B' + \\
& \quad B(\Gamma_j^+ Y + \Gamma_j^- Z) + 2\alpha_1 W < 0 \\
& WA' + AW + (Y'\Gamma_j^+ + Z'\Gamma_j^-)B' + \\
& \quad B(\Gamma_j^+ Y + \Gamma_j^- Z) + 2\alpha_2 W > 0
\end{aligned} \quad (22)$$

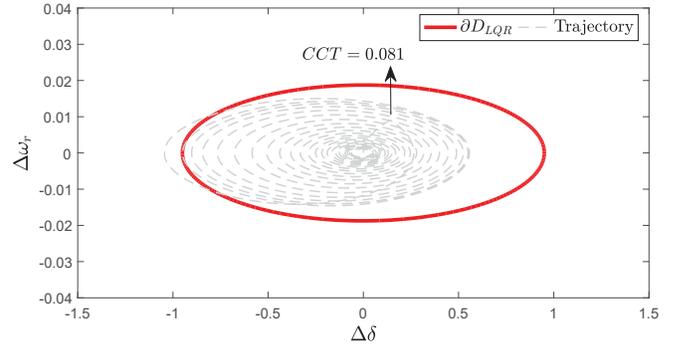

Fig. 3. Boundary of estimated DA and trajectory of nonlinear SMIB system with LQR controller and critical fault duration of $CCT = 0.081s$.

where controller matrix $F$ can be obtained from $F = YW^{-1}$. Matrices $\Gamma^+$ and $\Gamma^- \in \Re^{m \times m}$ are diagonal matrices whose diagonal elements take the value 1 or 0, and $\Gamma^- + \Gamma^+ = I_m$ where $j = 1, \ldots, 2^m$. For a single input system, these matrices can be either 0 or 1. The last two inequalities in optimization (22) are to restrict the pole placement region to a strip of the complex plane between $-\alpha_1$ and $-\alpha_2$ to avoid high gains in the controller.

**Example** (continued): Using the above optimization (22), a controller can be designed to enlarge the DA of the SMIB system. Assuming $\alpha_1 = 0$ and $\alpha_2 = 80$, the enlarged DA is obtained as $\mathcal{D}_{Enl} = \varepsilon(P, 1)$ with:

$$P = \begin{bmatrix} 0.8411 & 0.6754 & 0.6719 & 0.0015 \\ 0.6754 & 2180.8 & -27.7163 & -0.0454 \\ 0.6719 & -27.7163 & 0.9029 & 0.0018 \\ 0.0015 & -0.0454 & 0.0018 & 3.7605 \times 10^{-6} \end{bmatrix} \quad (23)$$

where the optimized state-feedback controller is:

$$F_{Enl} = \begin{bmatrix} -3.3026 & 98.2739 & -3.9459 & -0.0081 \end{bmatrix} \quad (24)$$

Fig. 4 illustrates the effectiveness of the proposed approach in designing the SDC, which results in a significantly larger DA toward the boundary of null controllable region $\partial\mathcal{C}$ for the closed loop system, in compared to the original LQR controller. Numerical simulations reveal that system with optimized state-feedback controller shows improvement in the $CCT$ to 0.109 s. Fig. 5 depicts a cut of the enlarged DA and extremal trajectory of the nonlinear SMIB system with optimized controller, which demonstrates the satisfactory accuracy.

Moreover, synchronizing and damping components of electrical torque can be used to compare the dynamic performance of the aforementioned controllers. The damping component (proportional to speed change) and the synchronizing component (proportional to angular change) are related to small-signal and transient stability, respectively. The corresponding coefficients are defined according to the following equation:

$$\Delta T_e = K_D \omega_s \Delta \omega_r + K_S \Delta \delta \quad (25)$$

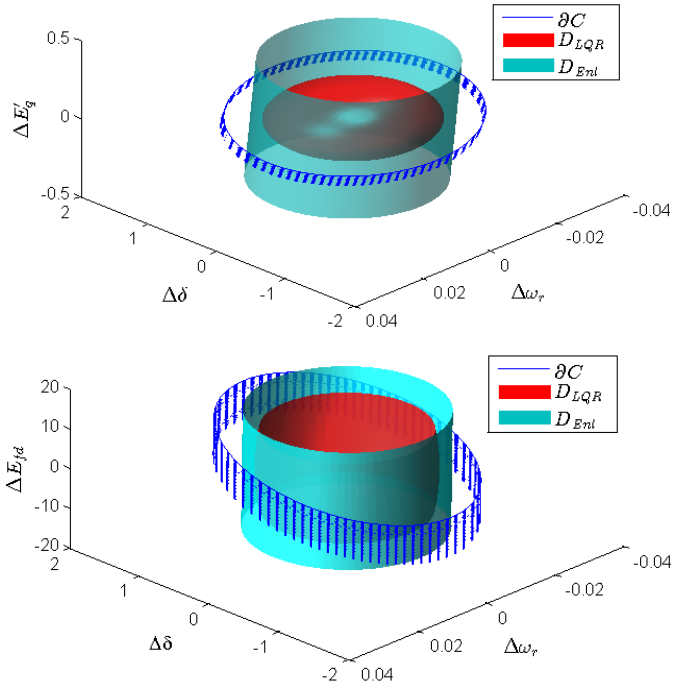

Fig. 4. Boundary of null controllable region ($\partial C$), estimated DA for LQR controller ($D_{LQR}$) and enlarged DA ($D_{Enl}$) in the presence of actuator saturation.

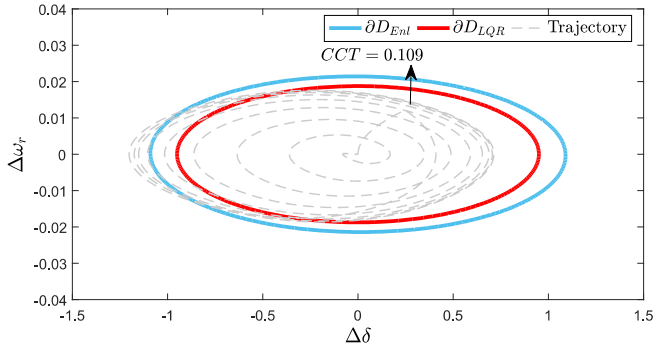

Fig. 5. Comparison of estimated DA and trajectory of nonlinear SMIB system with LQR controller and critical fault duration of $CCT = 0.109s$.

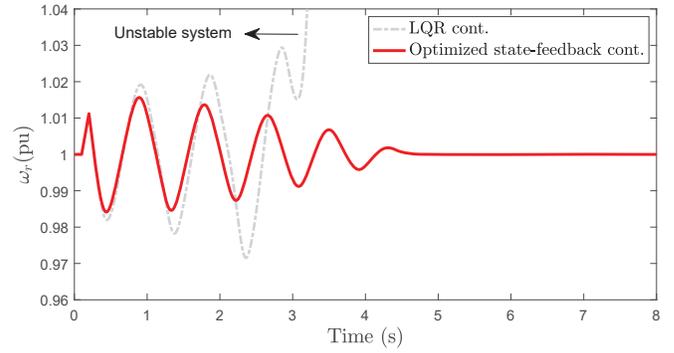

Fig. 6. Comparison of transient responses for the system with fault duration of $t_f = 0.1s$.

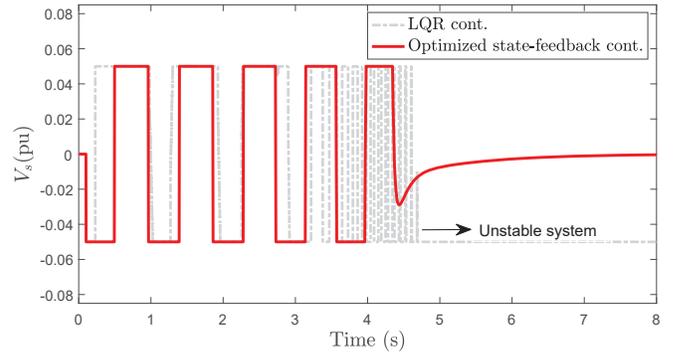

Fig. 7. Comparison of supplementary control signals for system with fault duration of $t_f = 0.1s$.

TABLE I
CCT AND COEFFICIENT COMPARISON FOR SMIB SYSTEM WITH DIFFERENT CONTROLLERS.

| SDC type | CCT | $K_D$ | $K_S$ |
|---|---|---|---|
| Without controller | 0.0 s | -0.00242 pu/(rad/s) | 1.0003 pu/rad |
| LQR controller | 0.081 s | 0.00223 pu/(rad/s) | 0.7078 pu/rad |
| Optimized state-feedback controller | 0.109 s | 0.00946 pu/(rad/s) | 0.8091 pu/rad |

where values of parameters $K_D$ and $K_S$ are estimated using the breaking algorithm [31] based on the angle, speed and torque response of the nonlinear system with actuator saturation. As shown in Table I, the system with optimized controller has larger CCT, higher damping ratio and synchronization coefficient. Fig. 6 shows a comparison of the transient response of the closed loop system for a fault duration of $t_f = 0.1s$ in the presence of actuator saturation. It can be seen that the system with LQR controller will be unstable. However, the optimized state-feedback controller significantly improves the damping of the closed loop system. Fig. 7 shows that the proposed controller uses the full feasible control range of $\left[V_s^{\min}, V_s^{\max}\right]$ to enlarge the DA and its performance is close to a bang-bang control law.

## V. CONCLUSIONS

In this paper, a new approach to design supplementary damping controllers by taking into account the effects of saturation limits is introduced. The problem of determining the optimal estimation of DA for SMIB power in the presence of saturation on the control signal is considered. To increase the region of stability, state-feedback controllers are designed to enlarge the guaranteed DA. Consequently, the enlargement of the DA of the post-fault system effectively increases the CCT, which is an important measure of transient stability. Detailed dynamic simulation results demonstrate that the proposed controllers use the available control range to effectively enlarge the DA, improve the damping and enhance the stability in the presence of hard saturation.

## Appendix A

$K$ constants in terms of system parameters can be summarized as follows:

$$\Delta_e = R_e^2 + (X_d' + X_e)(X_q + X_e)$$
$$K_1 = -\frac{1}{\Delta_e}\big[I_q V_\infty(X_d' - X_q)\big((X_q + X_e)\sin(\delta) - R_e\cos(\delta)\big)$$
$$+ V_\infty\{(X_d' - X_q)I_d - E_q'\}\{(X_d' + X_e)\cos(\delta) + R_e\sin(\delta)\}\big]$$
$$K_2 = \frac{1}{\Delta_e}\big[I_q\Delta_e - \big(I_q(X_q + X_e) + R_e I_d\big)(X_d' - X_q) + R_e E_q'\big]$$
$$K_3 = \frac{\Delta_e}{\Delta_e + (X_d - X_d')(X_q + X_e)}$$
$$K_4 = \frac{V_\infty}{\Delta_e}(X_d - X_d')\big[(X_q + X_e)\sin(\delta) - R_e\cos(\delta)\big]$$
$$K_5 = \frac{1}{\Delta_e V_t}\big[V_d X_q R_e V_\infty \sin(\delta) + V_d X_q V_\infty(X_d' + X_e)\cos(\delta)$$
$$+ V_q X_d' R_e V_\infty \cos(\delta) - V_q X_d' V_\infty(X_q + X_e)\sin(\delta)\big]$$
$$K_6 = \frac{1}{\Delta_e V_t}\big[V_d R_e X_q - V_d X_d'(X_q + X_e)\big] + \frac{V_d}{V_t}$$